\documentclass[fleqn,10pt]{wlscirep}
\usepackage[utf8]{inputenc}
\usepackage[T1]{fontenc}
\usepackage{lineno}
\usepackage{bm}

\title{Guided Diffusion for Fast Inverse Design of Density-based Mechanical Metamaterials}

\author[1,$\dag$]{Yanyan Yang}
\author[1,$\dag$]{Lili Wang}
\author[1,*]{Xiaoya Zhai}
\author[2]{Kai Chen}
\author[3]{Wenming Wu}
\author[1]{Yunkai Zhao}
\author[1]{Ligang Liu}
\author[1,*]{Xiao-Ming Fu}
\affil[1]{School of Mathematical Sciences, University of Science and Technology of China, Hefei, Anhui 230026, P.R. China}
\affil[2]{Beijing Academy of Artificial Intelligence, Beijing 100190, P.R. China}
\affil[3]{Hefei University of Technology, Hefei, Anhui 230026, P.R. China}

\affil[*]{Corresponding author(s): Xiaoya Zhai (xiaoyazhai@ustc.edu.cn); Xiao-Ming Fu (fuxm@ustc.edu.cn)}

\affil[$\dag$]{These authors contributed equally to this work}

\begin{abstract}
Mechanical metamaterials are synthetic materials that can possess extraordinary physical characteristics, such as abnormal elasticity, stiffness, and stability, by carefully designing their structure. 
The representation of metamaterials through high-resolution voxels has the potential to reveal delicate local structures with unique mechanical properties. However, this approach results in a substantial computational burden.
To this end, this paper proposes a fast inverse design method, whose core is an advanced deep generative AI algorithm, to generate voxel-based mechanical metamaterials.
 The dataset construction involves generating the initial microstructure with a resolution of $128^3$ through optimization and employing an active learning strategy to augment the data, thereby obtaining a dataset with extensive coverage of physical properties. 
 Specifically, we use a self-conditioned diffusion model.
 The model is capable of generating a microstructure to approach the specified homogenized tensor matrix in just 0.42 seconds.
 Accordingly, this rapid inverse design tool facilitates the exploration of extreme metamaterials, the sequence interpolation in metamaterials, and the generation of diverse microstructures for multiscale design.
 This flexible and adaptive generative tool is of great value in structural engineering or other mechanical systems and can stimulate more subsequent research.
\end{abstract}
\begin{document}

\flushbottom
\maketitle

\thispagestyle{empty}


\section*{Introduction}



Architectured metamaterials, a novel frontier in materials research, have opened up new possibilities for creating materials with unique and unconventional functionalities. They offer precise control over a range of physical properties, including mechanical strength~\cite{meza2015resilient,jiang2016highly,surjadi2019mechanical}, Poisson's ratio~\cite{chen2017lattice,li2017harnessing,khan2021novel,farzaneh2022sequential,zhang2023self}, and bulking behavior~\cite{ghaedizadeh2016geometric,yang2019multi,yves2017crystalline}. The surveys~\cite{bertoldi2017flexible,yu2018mechanical,jiao2023mechanical}  
 provide a comprehensive overview of the research progress in mechanical metamaterials.

Once the properties of the base material are given, the mechanical properties of the metamaterial are determined by its geometry.  
Mathematically, geometry can be expressed in various forms, such as parametric functions, implicit functions, discrete meshes, and voxel representations.
 Voxel representation has two main advantages compared with other representations: (1) it is flexible for representing various geometries, thus possessing various mechanical properties, and (2) it is regular and simple, reducing the design difficulty of the metamaterial generation algorithm. 
Moreover, high-resolution metamaterials represented by a large number of voxels can have delicate local structures, thus showing exciting mechanical properties.


Voxel-based metamaterials are generally obtained using the topology optimization method to solve the inverse homogenization problem~\cite{hassani1998review}. 
 However, this method has three limitations due to the large design space, which is based on density.
First, high-resolution topology optimization still requires expensive computing resources and considerable computing time, although various excellent algorithms have been proposed, such as the multi-CPU framework~\cite{aage2015topology,borrvall2001large,aage2017giga}, GPU computation~\cite{challis2014high,xia2017gpu,wu2015system,zhang2023optimized}, and adaptive mesh refinement~\cite{stainko2006adaptive,wang2010dynamic}.
Second, since the optimization problem is highly nonlinear and nonconvex, the optimized metamaterials are closely related to the initialization. In practice, there is almost no rule for choosing initial metamaterials to be optimized with high stiffness (or even close to extreme values), and we can only perform tedious trials and errors.
 Third, due to the high computational burden and the dependence on initial metamaterials, diverse generated metamaterials have poor geometric connectivity, and it is challenging to create metamaterial sequences with extreme mechanical properties, thereby decreasing their performance in multiscale design~\cite{du2018connecting,garner2019compatibility}.

The data-driven inverse design of mechanical metamaterials is a possible way to address these limitations~\cite{lee2023data}.
It can inversely design mechanical metamaterials~\cite{zeng2023deep,zheng2023deep,zheng2021controllable,kumar2020inverse} and quickly assemble and generate multiscale systems~\cite{patel2022improving,seo2023dl}.
Specifically, since the goal of a deep generative AI model is highly similar to that of an inverse design algorithm, which is to generate a microstructure meeting specific requirements, deep generative AI models have been widely used to create microstructures~\cite{jiao2021artificial,song2023artificial}.
In addition to the representation of models, datasets play a crucial role in this kind of algorithm.
The existing microstructure datasets mainly include three types: 
(1) low-resolution datasets using pixel/voxel representation~\cite{wang2020data, chan2021metaset,10.1115/1.4055925}, 
(2) datasets with varying geometric parameters such as the width of the truss ~\cite{korshunova2021uncertainty, bastek2022inverting, zheng2023unifying}, and 
(3) implicit surface-based datasets that mainly contain triply periodic minimal surfaces (TPMS)~\cite{wang2022ih} and spinodoid surfaces~\cite{kumar2020inverse}. 
Unfortunately, the current datasets exhibit limitations.
For instance, many databases are configured in either 2D pixel or 3D voxel structures with low resolutions, which significantly restricts the performance of these datasets. While structures represented by geometric parameters have fewer design variables, their coverage of performance space is also limited. The implicit representation of datasets belongs to a specific function class, which is more suitable for certain scenarios; for example, the spinodoid structure dataset is ideal for bionic bone structure optimization.
Therefore, the need for advanced data-driven methods and the generation of high-resolution voxel-based datasets with high property coverage to explore mechanical metamaterials is becoming increasingly urgent.

We construct a voxel-represented mechanical metamaterial dataset at a high resolution of $128^3$, with partial coverage of modulus variations.
Each metamaterial is obtained using the LIVE3D framework~\cite{zhang2023optimized}, which requires only a GPU.
Specifically, the bulk modulus, shear modulus, and Poisson's ratio are the optimization objectives, and the constrained volume fraction ranges from 0.2 to 0.9. 
Furthermore, isotropic constraints are added to optimize part of the microstructures.
Thus, the optimized dataset includes both isotropic and anisotropic metamaterials. 

\begin{figure}[!t]
    \centering
    \includegraphics[scale=0.34]{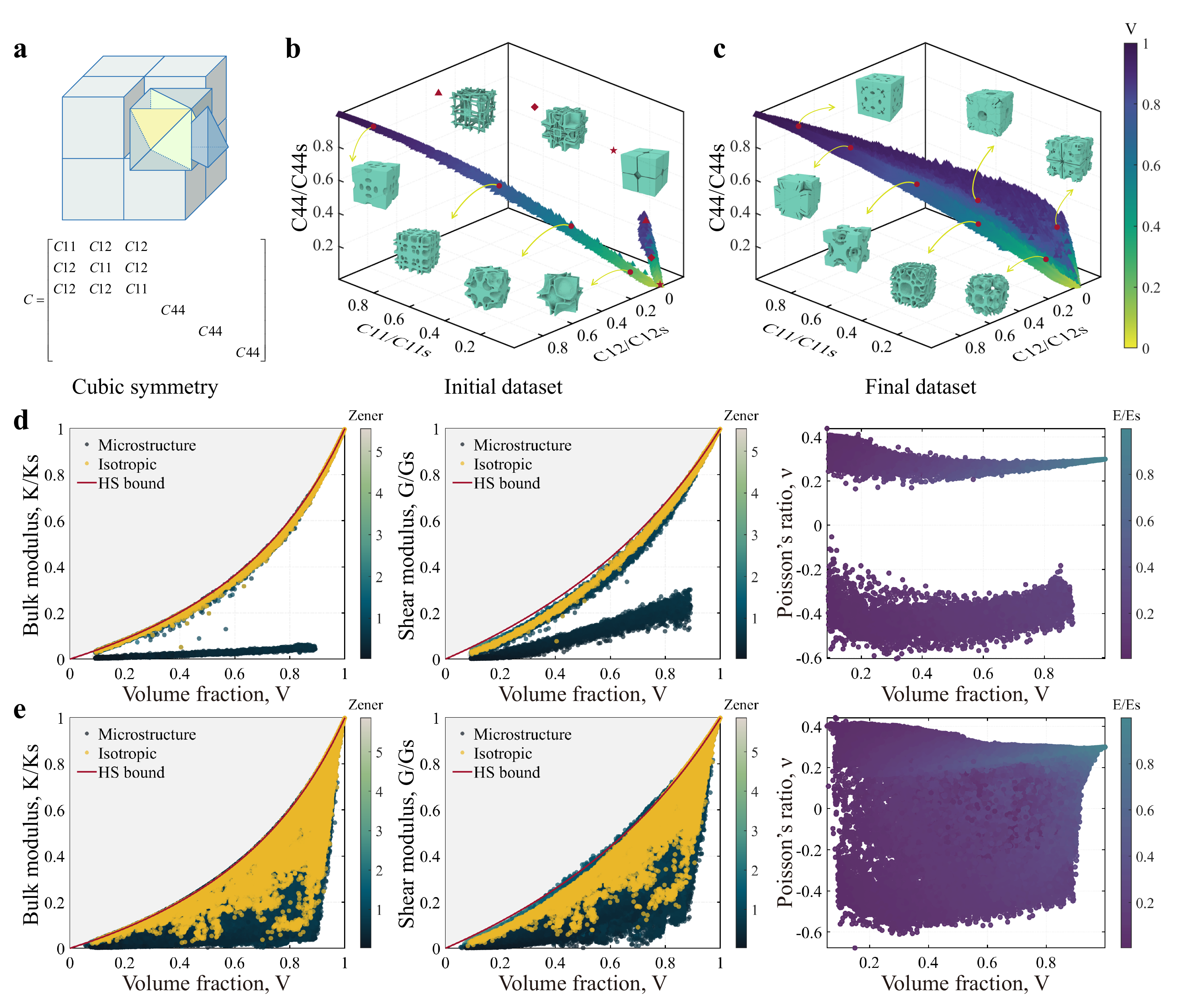}
    \vspace{-3mm}
    \caption{\textbf{Overview of the voxel-based metamaterial dataset. }
    \textbf{a} Cubic symmetry and elasticity tensor. 
    \textbf{b} Initial dataset.
    \textbf{c} Final dataset.
    \textbf{d} Bulk Modulus, shear Modulus, and Poisson's ratio of the initial dataset. 
    \textbf{e} Bulk Modulus, shear Modulus, and Poisson's ratio of the final dataset.
    }
    \label{fig:dataset}
\end{figure}

This paper introduces a fast inverse design method to generate voxel-based mechanical metamaterials.
Central to the technique is the adoption of the self-conditioned diffusion model (an advanced deep generative AI algorithm), which produces a microstructure with a resolution of $128^3$ to approach the specified homogenized elasticity tensor matrix in just 0.42 seconds on an NVIDIA GeForce RTX 3090 GPU, substantially enhancing the efficiency of the microstructure inverse design. 
Since the initial dataset only partially covers modulus variations, we apply an active learning technique to train the diffusion model and augment the dataset alternately. This results in a dataset featuring more diverse shapes, broader coverage of modulus values, and a diffusion model with higher generation accuracy (Figure~\ref{fig:dataset}). 

\begin{figure}[t]
    \centering
    \includegraphics[scale=0.33]{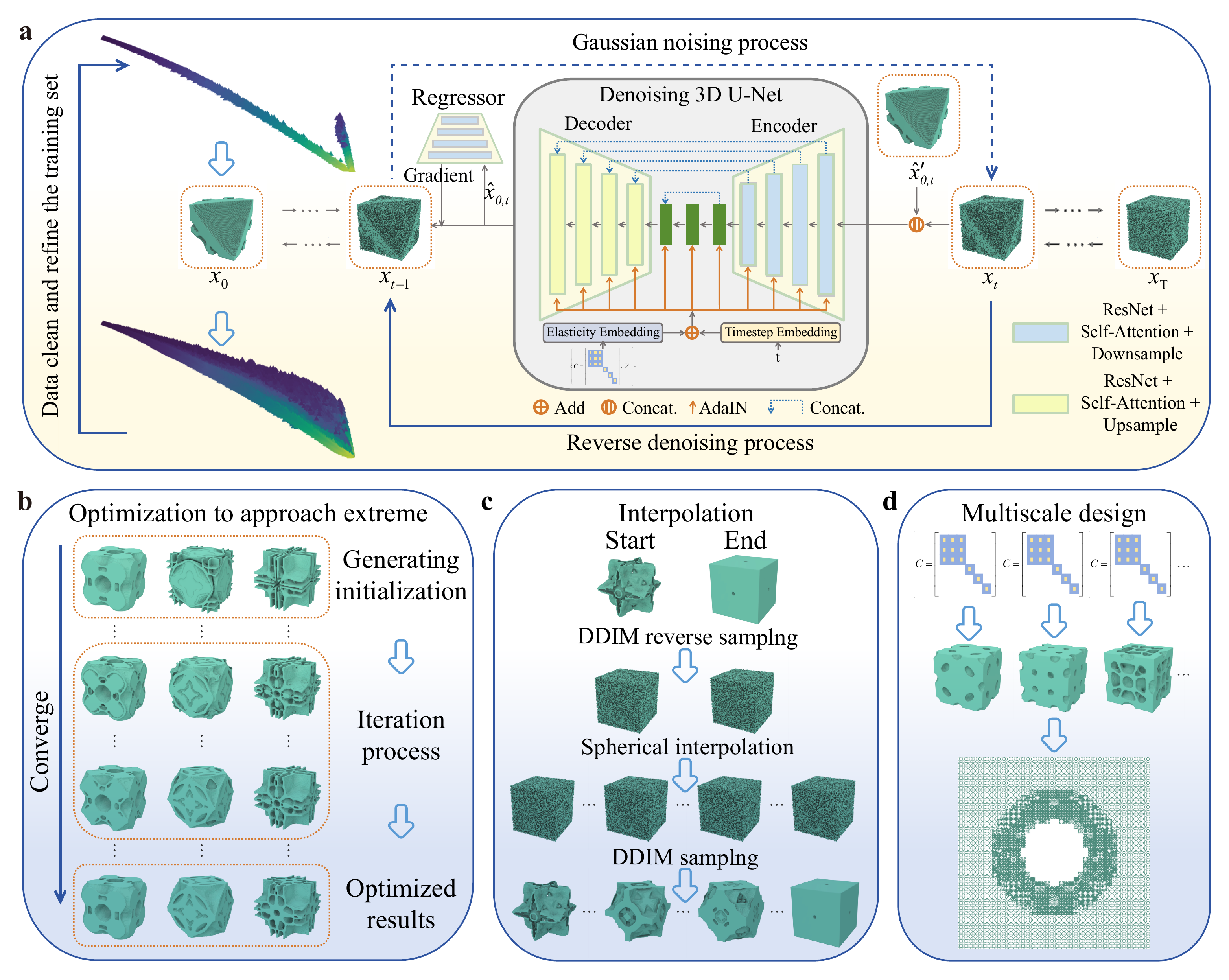}
    \caption{
   \textbf{Generative model framework and its three applications.}
    \textbf{a} Framework of the diffusion model. 
    \textbf{b} Using generated results as initials to do optimization approaching extreme. 
    \textbf{c} Properties and shape interpolation of microstructure sequences based on the denoising diffusion implicit model (DDIM)~\cite{song2020denoising}. 
    \textbf{d} Multiscale design.}
    \label{fig:network}
\end{figure}

Consequently, the model is successfully used for efficient inverse design, facilitating the exploration of extreme metamaterials, sequence interpolation in metamaterials, and the generation of diverse microstructures (Figure~\ref{fig:network}). 
First, the diffusion model has been proven to be effective in generating initial microstructures for subsequent traditional topology optimization. For example, we use the diffusion model to construct an initial microstructure with a negative Poisson’s ratio of $-0.54$ and then realize the final extreme metamaterial with a negative Poisson's ratio of $-0.63$ through subsequent topology optimization.
 Second, according to the interpolation capability of the diffusion model, we generate a series of microstructure sequences that closely approach the theoretical upper limit. 
 Finally, since the diffusion model can generate geometrically diverse microstructures, it is more likely to have high geometric connectivity while approaching the target properties in multiscale design.


\section*{Results}

\subsection*{Initial metamaterial dataset}

We construct a large-scale voxel-based metamaterial dataset under cubic symmetry constraints (Figure~\ref{fig:dataset}\textbf{a}).
To create a sufficiently large design space for a voxel-represented metamaterial dataset, we establish microstructures at a resolution of $128^3$ and subsequently store $\frac{1}{8}$-th of them due to the cubic symmetry constraint, i.e., $64^3$ microstructures.
The representation is the basis for assembling a dataset incorporating a sizable assortment of voxel-based lattices, effectively capturing a broad range of mechanical properties. 
To construct the initial metamaterial dataset, we use the LIVE3D framework~\cite{zhang2023optimized} to optimize a collection of metamaterials (Figure~\ref{fig:dataset}\textbf{b}). 
Specifically, trigonometric functions are adopted to cover various initial density fields. 
The basis functions of the initialization are $\{\cos 2\pi k\bm{x}, \sin 2\pi k\bm{x}\}$ $(0<k\leq n)$, where $\bm{x} \in \mathbf{R}^3$ is the coordinate of the element's center. 
We formulate the optimization model with the modulus as the objective function and the volume constraint.
Consequently, the constructed dataset has partial coverage of modulus variations.

\subsection*{Self-conditioned diffusion model}

\subsubsection*{Model illustration}
As described in Denoising Diffusion Probabilistic Models (DDPM)~\cite{ho2020denoising}, a diffusion model comprises both forward and reverse processes. 
In the forward process, Gaussian noise is incrementally introduced to the data point $x_0$ until it evolves into pure Gaussian noise, denoted as $x_T$, which can be defined as follows:
\begin{equation}
    x_t = \sqrt{\gamma_t}x_0+\sqrt{1-\gamma_t}\epsilon,
\end{equation}
where $\epsilon\sim\mathcal{N}(0,I)$ and  $\gamma_t$ monotonically decreases from 1 to 0.
The reverse process restores Gaussian noise $x_t$ to a data point $x_0$ by denoising it step by step. 
At each step, the prediction of $x_{t-1}$ from $x_t$ is facilitated by a neural network, typically a U-Net, which predicts $x_{t-1}$, $\epsilon$, or $x_0$. 
In our work, we use a U-Net $f(x_t, \hat{x}'_0, t, \text{conditions})$ to predict $x_0$, where $\hat{x}'_0$ is an estimation of $x_0$ from the previous prediction, which is introduced by
self-conditioning technique~\cite{chen2022analog}.
During training,  $\hat{x}'_0$ is set to $f(x_t,0,t, \text{conditions})$ with 50\% probability; otherwise, it is set to 0.
The conditions include the elasticity tensor ($C_{11}, C_{12}, C_{44}$) and volume fraction. 
We first normalize $C_{11}, C_{12}, $ and  $ C_{44}$ into $[0,1]$ separately. 
Then, the condition's four components are encoded by learnable sinusoidal embeddings and integrated into the model through classifier-free guidance~\cite{ho2022classifier}. 
By setting them to a predetermined special value (we use -1 in the experiments) separately or simultaneously, we can generate structures conditioned solely on either the elasticity tensor or the volume fraction or without conditions.
The loss function is:
\begin{equation}
    \mathcal{L}_{x_0}=E_{\epsilon\sim\mathcal{N}(0,I),t\sim U(0,I)}||\hat{x}_0-x_0||_2^2,
\end{equation}
where
\begin{equation}
    \hat{x}_0 = f(x_t,\hat{x}_0',t, \text{conditions}).
\end{equation}
Since the data are cubic symmetric, we force the results to be cubic symmetric at the end of the forward process, significantly improving the network performance.

\subsubsection*{Active learning strategy}

As shown in Figure~\ref{fig:dataset}, the initial dataset consists of only microstructures with either a high modulus or low Poisson’s ratio. Consequently, the coverage range is limited.
Previous methods~\cite{zhu2017two,wang2020data} use perturbations to extrapolate substantial datasets from limited datasets. 
We also try a three-dimensional perturbation method, yet the outcomes fall short of the ideal.
Specifically, the surfaces of the structures derived through perturbation consistently exhibit characteristics reminiscent of potholes. A detailed discussion is given in Supplementary Figure 1.

Instead, we propose the use of active learning to augment the dataset.
In essence, active learning means repeatedly taking the results generated by the previous model and putting them back into the training set for further training.
With each iteration, the data become richer.
Due to the high fidelity of the generative model, it maintains its quality. The improved dataset results in higher accuracy and an expanded generative scope for the model. 
Then, in the subsequent iteration, the more powerful generative model will produce even richer data.

The generative model may occasionally generate unreliable results. If these are left in the training set for further training,  the model's reliability will significantly diminish. 
Since we are generating one-eighth of unit cells, there are two potential types of unreliable generation results: (1) lack of connectivity, leading to disconnected unit cells, and (2) absence of voxels at the boundary, leading to either disconnected 
or lacking voxels at the boundary, which violates the assumptions of homogenization theory. 
After removing these two types of undesirable results, further data cleaning is necessary. 
We remove excessively dense data points in the $C_{11}-C_{12}-C_{44}-V$ space to ensure a uniform data distribution, thus preventing bias in the model.

After undergoing three cycles of generation, cleaning, data augmentation, and retraining, we eventually obtain a comprehensive dataset that looks like the convex hull of the initial dataset (Figure~\ref{fig:dataset}). 
Further analysis of the iteration process is provided in Supplementary Figure 2.

\subsubsection*{Performance of the diffusion model}

\paragraph{Fast inverse design} 
As the voxel resolution progressively increases, the computational costs will correspondingly increase via traditional topology optimization methods.
In essence, computer memory and computing time are enormous
challenges.
After fully using the hardware resources, the inverse design problem is solved by the multi-CPU framework~\cite{aage2015topology} and GPU computations~\cite{zhang2023optimized}. 
Nonetheless, conventional approaches typically require several hours for resolution. 
Our guided diffusion model for attaining one microstructure with a resolution of $128^3$ requires only 0.42 seconds under a specified elasticity tensor.
\begin{figure}[!t]
    \centering
    \includegraphics[scale=0.12]{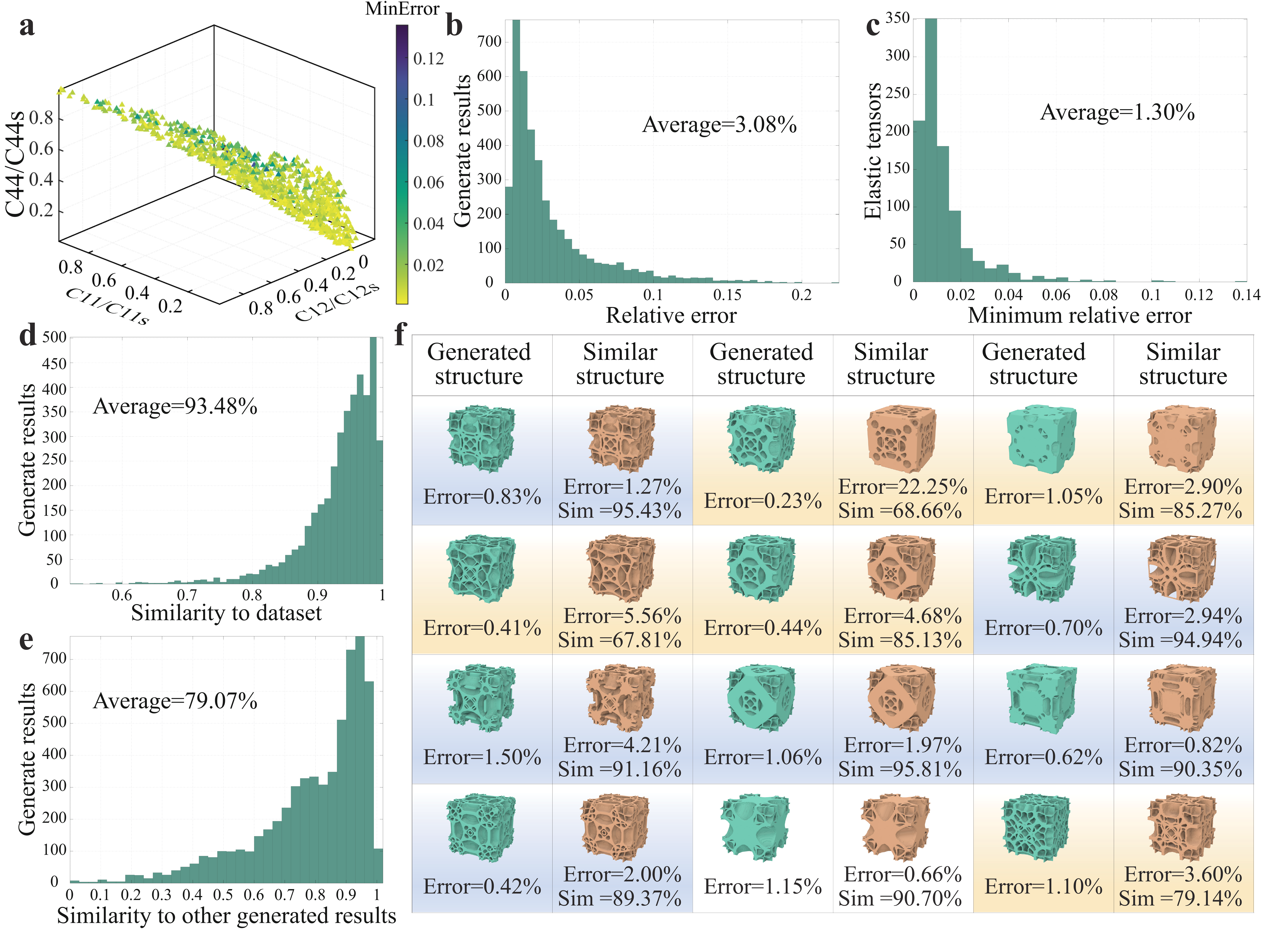}
    \caption{\textbf{Performance of the diffusion model.} 
    \textbf{a-c} \textbf{Generalization and accuracy.} We randomly took 1000 elasticity tensors in coverage of the dataset and generated four structures for each elasticity tensor. 
    \textbf{a} Minimum relative error of the four generated results for each given elasticity tensor.
    \textbf{b}  Histograms of the relative errors of the 4000 generated results. \textbf{c} Histograms of the minimal relative errors of the 1000 given elasticity tensors.
    \textbf{d-f} \text{Diversity and novelty.} We describe novelty by the similarity between the generated results and the most similar structure in the dataset and show diversity by the similarity between the four structures generated by the same property condition. 
    \textbf{f} We show three sets of structures (4 structures generated by the same condition and their most similar structures in the dataset) to demonstrate diversity and novelty further. 
We unexpectedly observe that the generated structures are mostly closer to the property conditions than their most similar structure in the dataset. 
    Light yellow boxes mark the examples where the shape changes significantly and the accuracy improves. 
    Light cyan boxes mark the examples where the shape is almost unchanged and the accuracy improves. 
    This shows that our model does learn the relationship between shapes and properties rather than remember the dataset.
    }
    \label{fig:generation}
\end{figure}
In the following two paragraphs, we analyze the generation performance conditioned on an elasticity tensor. A similar analysis for generation conditioned on the elasticity tensor and volume fraction, or solely on the volume fraction, is provided in Supplementary Figure 3.
\paragraph{Accuracy and diversity}
Our primary goal is to generate diverse microstructures with prescribed elasticity tensors.
For accuracy assessment, we measure the relative error of the elasticity tensor $\mathbf{C}$ as the average of the relative errors associated with its three independent components, namely, $C_{11}$, $C_{12}$, and $C_{44}$. The error statistics are as follows:
\begin{equation}
  \centering
  Error(\mathbf{C}^\text{cond}, \mathbf{C}^\text{gen}) = \frac{1}{3}\left(\left|\frac{C_{11}^\text{cond}-C_{11}^\text{gen}}{C_{11}^{\max}-C_{11}^{\min}}\right|+\left|\frac{C_{12}^\text{cond}-C_{12}^\text{gen}}{C_{12}^{\max}-C_{12}^{\min}}\right|+\left|\frac{C_{44}^\text{cond}-C_{44}^\text{gen}}{C_{44}^{\max}-C_{44}^{\min}}\right|\right), 
  \label{eq:error}
\end{equation}
where $\mathbf{C}^\text{cond}$ is the target elasticity tensor as the condition of the diffusion model and $\mathbf{C}^\text{gen}$ is the elasticity tensor of the generated structure. 
Figure~\ref{fig:generation}b and Figure~\ref{fig:generation}c show the generation accuracy on the randomly sampled test set of the elasticity tensor, with an average error of $3.08\%$, which is further reduced to $1.30\%$ if we choose the best structure among the four generated structures corresponding to each target tensor. 
Among all 4000 generated results, we found that all structures had voxels on the boundary, and there were 106 disconnected structures. These disconnected structures had only a small volume fraction. For these patients, we included the largest connected component in the analysis.

Our diffusion model can generate many differently shaped microstructures for the same target elastic tensor. 
For the statistical analysis of diversity (and later novelty), we define the shape similarity of two microstructures of $S_1$ and $S_2$: 
\begin{equation}
 \centering
    Similarity(S_1, S_2) = \frac{\#(\text{same voxels for } S_1 \text{ and } S_2)} {\sqrt{\#(\text{voxels in }S_1) * \#(\text{voxels in }S_2)}}, 
  \label{eq:sim}
\end{equation}
where $\#$ is the number. 
In Figure~\ref{fig:generation}e, we compute the pair-to-pair similarity between the four structures generated by each target tensor, and the results further show that we yield very different microstructures for the same target tensor. 

In classifier-free guidance, we can make a trade-off between accuracy and diversity by adjusting the guidance scale. In all the experiments shown in Figure~\ref{fig:generation}, we set the guidance scale to 1. More analysis of the guidance scale can be found in Supplementary Table 1.
Furthermore, training an additional regressor is required only to enable generation under any other conditions.  
In our interpolation task (Figure~\ref{fig:interpolation}), we use a regressor to predict the ratio between the bulk modulus $K$ and the upper limit of the bulk modulus $K_b$ under the same volume fraction, with a coefficient of
determination $R^2=0.99$.

\paragraph{Novelty and generalizability} 
Additionally, our goal is not only for the model to memorize the training set but also to genuinely comprehend the relationships between the shape and properties of the microstructures. 
To assess the novelty, we continue to utilize the previously defined shape similarity in Eq.~\eqref{eq:sim} and further define the similarity between a microstructure and a dataset as the maximum similarity between that structure and all microstructures in the dataset. 
Figure~\ref{fig:generation}d illustrates the similarity between the generated results and the training set, providing evidence of our model's ability to generate novel microstructures. 
In Figure~\ref{fig:generation}f, we further illustrate three sets of results, comparing the generated outcomes with the microstructures in the dataset that exhibit the highest similarity. 
We are pleasantly surprised that most generated results are closer to the target property than their most similar microstructures in the dataset. 
Some exhibited noticeable shape changes, while others achieve greater proximity to the target property through subtle alterations in shape. 
This suggests that our model has indeed learned the relationship between the shapes of microstructures and their properties. 
It not only provides powerful generative capabilities but also holds the potential to assist researchers in investigating the mechanisms behind various physical properties in the future.


\begin{figure}[!t]
    \centering
    \includegraphics[scale=0.116]{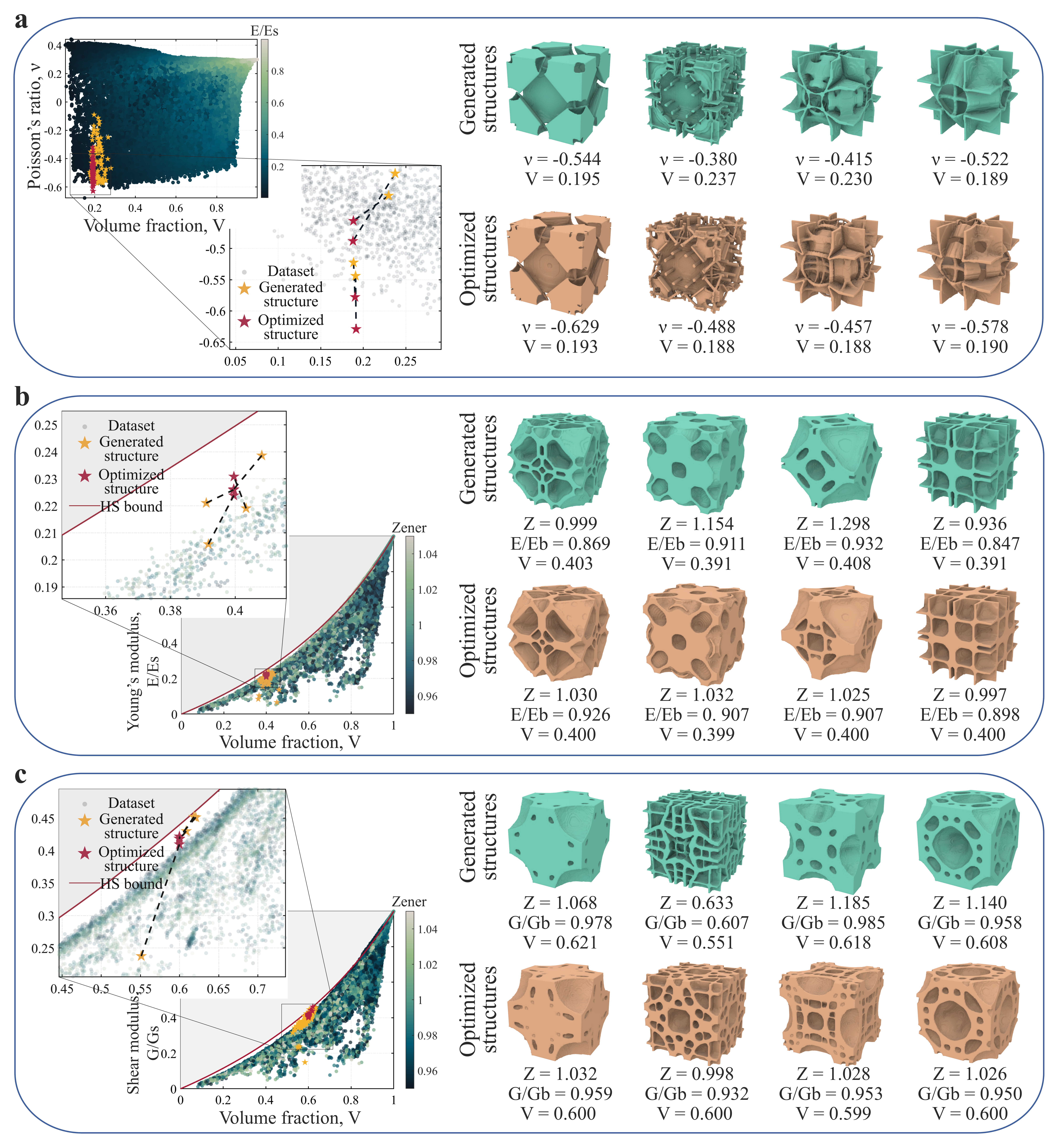}
    \caption{\textbf{Optimization with the generated structures as the initializations.} 
    For three optimization examples (\textbf{a} minimizing Poisson's ratio, \textbf{b} maximizing Young's modulus under isotropy, and \textbf{c} maximizing shear modulus under isotropy), we employ our model to generate initializations to help the optimization algorithm obtain extremer and more diverse microstructures. 
    In each example, we select 10 microstructures close to the target in the dataset and use their properties as the generation condition to generate 100 microstructures as the initial for optimization, of which 4 structures are shown on the right side, and their properties are shown in the zooming graph.
    }
    \label{fig:extreme}
\end{figure}

\subsection*{Gradient-based optimization to approach extremum}
Topology optimization using high-resolution frameworks has proven beneficial for exploring a broad range of metamaterials. However, it is a gradient-based optimization method whose results are closely related to the initial inputs. In other words, distinct initial values impact both the topologies and the performance of the final results. The uncertainty associated with the initial structure requires repeating the optimization process multiple times. The proposed diffusion model can provide better initial candidates for topology optimization algorithms. 

Figure~\ref{fig:extreme} illustrates three cases of the optimal design of voxel-based microstructures for extreme properties. 
In each of the three instances, we initially select ten microstructures from the dataset with properties close to the extreme and use their properties as conditions. 
We generate ten microstructures for each condition.
Our generative model produces a diverse set of initial structures near the target attributes, some of which even exceed the range of the dataset. 
This provides ample favorable starting points for voxel-based optimization algorithms, enabling the creation of a significant number of extreme metamaterials. 

In the first example, we optimize for the negative Poisson's ratio at a volume fraction of 0.2. 
The volume fraction and Poisson's ratio of the conditions of the diffusion model, the 100 generated results, and their corresponding optimization results are shown in Figure~\ref{fig:extreme}(a,left). 
Four sets of these microstructures are shown on the right side of Figure~\ref{fig:extreme}a. 
We successfully optimize for a negative Poisson's ratio of -0.629, which is significantly lower than the dataset's minimum value of -0.560. 
Our dataset includes data from optimizing for a negative Poisson’s ratio with random initializations. This comparison indicates that the initializations generated by our model outperform random initializations. 
In the second example (Figure~\ref{fig:extreme}b), the Young’s modulus is optimized under the constraints of isotropy and a volume fraction of 0.4. 
%
Moreover, in the third example (Figure~\ref{fig:extreme}c),  the shear modulus is optimized under the constraints of isotropy and a volume fraction of 0.6.
In these two examples, we first filter out microstructures in the dataset that are close to isotropic (with a Zener ratio greater than 0.95 and less than 1.05). 
Subsequently, we choose microstructures near the corresponding volume fraction and close to the Hashin–Shtrikman (HS) bound~\cite{hashin1962some} using its elasticity tensor as the generation condition. 
We achieve structurally rich, isotropic microstructures with performances close to the HS bound.

This generation-optimization approach can become a universal design process that is not limited by optimization goals and constraints. 
We obtain richer extreme structures by selecting more generation conditions for generating more microstructures as initial points for optimization. 
Additionally, the denoising diffusion implicit model (DDIM)~\cite{song2020denoising} is a deterministic sampling process, it allows reverse samplers. 
Thus, we can infer noise from the optimized structure and make minor perturbations to the noise to generate a more diverse set of extreme structures.

\begin{figure}[!t]
    \centering
    \includegraphics[scale=0.116]{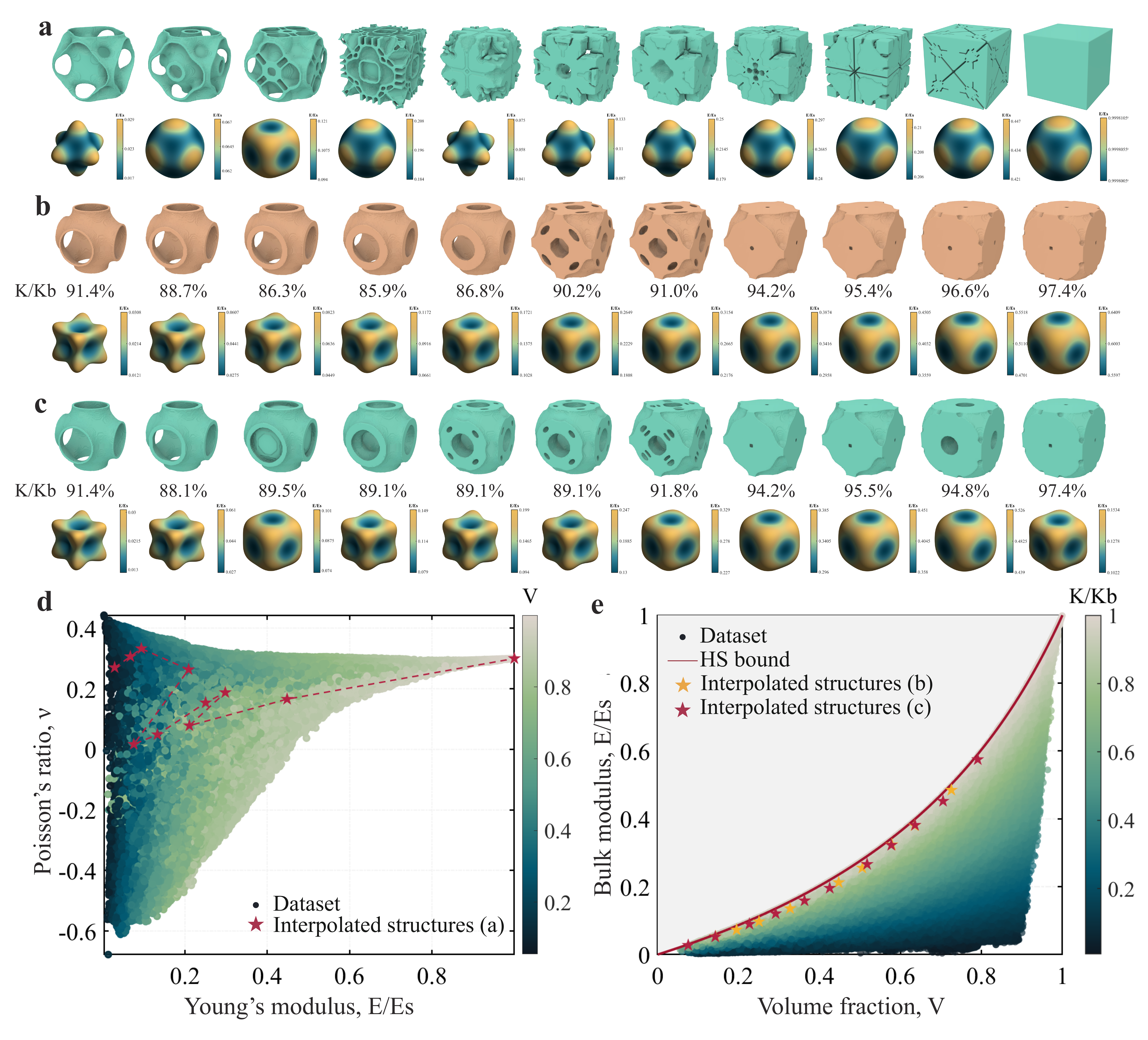}
    \caption{\textbf{Microstructure interpolation and microstructure family of extreme property.} \textbf{a.} Interpolation sequence with structures having extremely low and extremely high Young's modulus as the start and end points. \textbf{b,c.} Interpolation sequence with high bulk modulus. In \textbf{b}, only noise interpolation is performed, while in \textbf{c}, universal guidance is added. \textbf{d.} Interpolation trajectory in property space for \textbf{a}. \textbf{e.} Interpolation trajectory in property space for \textbf{b,c}.}
    \label{fig:interpolation}
\end{figure}

\subsection*{Microstructure interpolation}
Metamaterials typically take shape through the periodic or hierarchical arrangement of individual cells. These materials must exhibit geometric connectivity~\cite{du2018connecting} and maintain a continuous set of physical properties~\cite{garner2019compatibility}. 
Direct generation of sequential microstructures using topology optimization methods requires extra connectivity constraints. 
Interpolation~\cite{dhariwal2021diffusion} is a common application in generative models and is particularly well-suited for generating sequential microstructures.

During sampling, we start with Gaussian noise and progressively denoise to obtain a clear structure. 
By reversing this process, we can deduce the corresponding noise from a clear structure. 
From this perspective, noise can be considered a latent space for microstructures. 
We perform spherical interpolation on the noise to obtain a series of continuously varying microstructures. 
Specifically, given the start and end structure for interpolation, we first determine their corresponding Gaussian noises, denoted as $x_1$ and $x_2$. 
Subsequently, we generate a series of Gaussian noises $cos(\theta)x_0+sin(\theta)x_1$ where $\theta$ sweeps from $0$ to $\frac{\pi}{2}$. 
By using this series of noises for unconditional generation, we obtain a set of microstructures that smoothly change in both geometry and physical properties.

Figure~\ref{fig:interpolation}a shows an interpolation application case with start and end having extreme Young's moduli, and the corresponding elastic surfaces are shown in Figure~\ref{fig:interpolation}d. For such an extreme start and end, a smoothly transitioned sequence can be obtained, demonstrating the robustness of the interpolation method. Although we have not introduced physical information into the unconditional generative model, the noise of smooth transition leads to geometric shapes of smooth transition, resulting in their modulus also being smoothly transitional. However, such an interpolation solely for noise is not controllable for interpolation trajectories in the property space. This is consistent with past results~\cite{zheng2023unifying}. As a result, there has always been a gap between microstructure interpolation and multiscale design.

As a further effort, we slightly modify the interpolation algorithm and obtained a family of microstructures with bulk moduli close to the HS bound and volume fractions ranging from 0.07 to 0.8, as shown in Figure~\ref{fig:interpolation}c and Figure~\ref{fig:interpolation}e. 
We first obtain a series of microstructures with relatively high bulk moduli by selecting the start and end microstructures with bulk modulus values close to the HS bound, as shown in Figure~\ref{fig:interpolation}b. 
However, the bulk modulus of several microstructures falls below 86\% of the HS bound. 
Then, following the so-called universal guidance~\cite{bansal2023universal}, we add two controls for high bulk modulus and boundary similarity during the sequence generation and obtained the sequence shown in Figure~\ref{fig:interpolation}c. 
Its bulk modulus and transition smoothness are superior to those in Figure~\ref{fig:interpolation}b.

\begin{figure}[!t]
    \centering
    \includegraphics[scale=0.115]{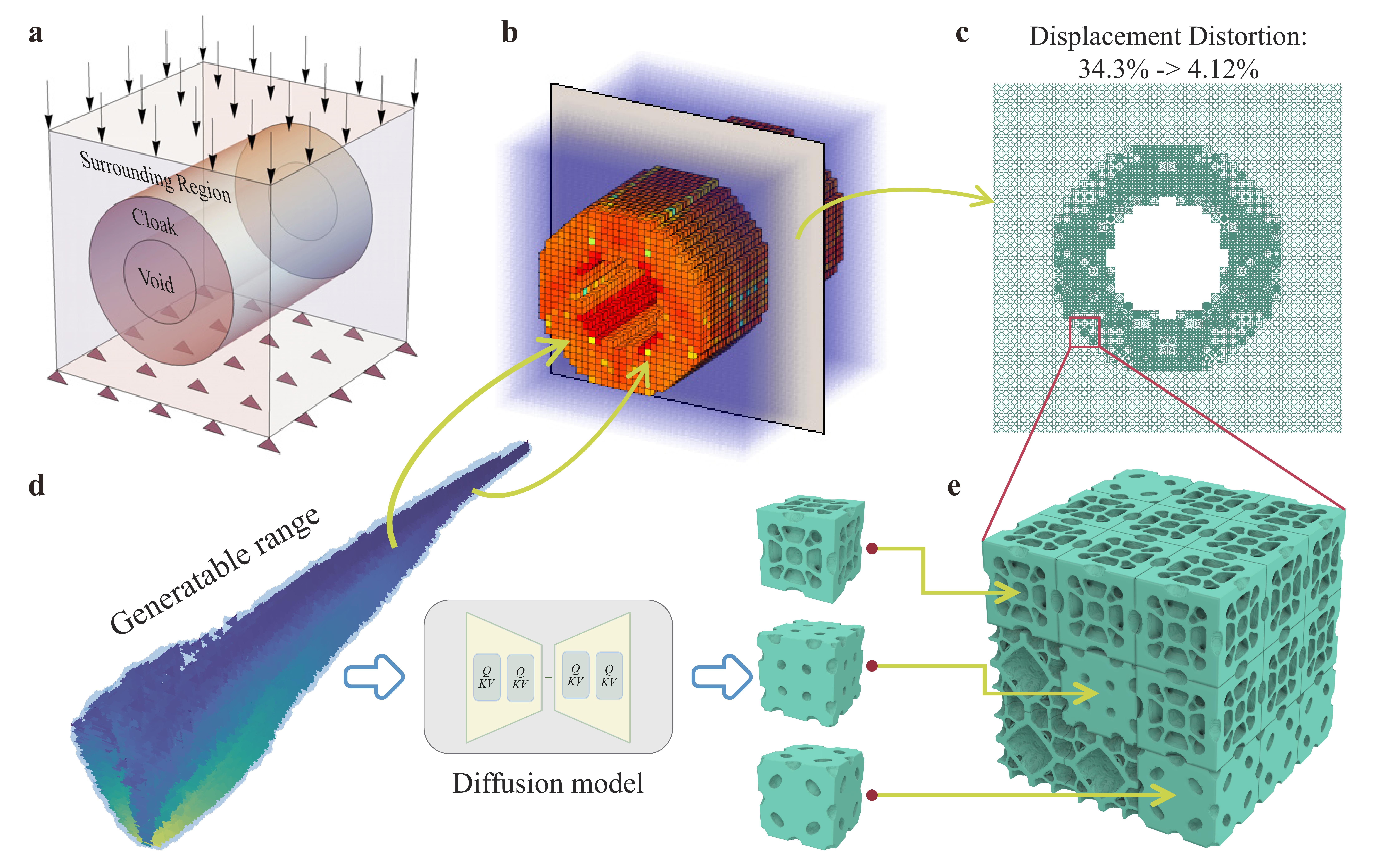}
    \caption{\textbf{Mechanical cloak via multiscale design.} 
\textbf{a} Problem definition. 
\textbf{b} C11 distribution of optimized result. 
\textbf{c} A cross-section of the optimized result. 
\textbf{d} The range of elasticity tensors that can be generated by the generative model. 
\textbf{e}  A 3*3 part of the optimized result. 
    }
    \label{fig:multiscale}
\end{figure}

\subsection*{Multiscale design}
We also employ our generative model for multiscale design. By trivially extending the optimization problem and methods for 2D mechanical cloaks~\cite{wang2022mechanical} to 3D, we design mechanical cloaks in 3D.  
The reference structures (absent voids and cloaks) consist of $40\times 40 \times 40$ periodically tessellated base cells selected from the metamaterial dataset. We design an invisibility cloak for such a situation by excising a cylinder with a radius of $\frac{20}{3}$ and subsequently filling the outer circular region (comprising a cylinder with an outer radius of $\frac{40}{3}$) with material to design the invisibility cloak.
Our surrounding structure is an isotropic structure identified from the dataset to emulate materials with isotropic properties, such as soil. 
Our designed invisibility cloak can reduce displacement distortion from 34.3\% to 4.12\%, achieving effective cloaking performance.

\section*{Discussion}
This paper introduces a fast inverse design method via an advanced deep generative AI algorithm to generate metamaterials matching a specified tensor. 
A high-resolution voxel-based dataset is generated through optimization and active learning methods. Its broad coverage facilitates a comprehensive exploration of mechanical metamaterials.
The self-conditioned diffusion model, the heart of our approach, is a powerful tool that can rapidly generate voxel-based mechanical metamaterials with a resolution as high as $128^3$. Its exceptional ability to approach the specified homogenized tensor matrix in just 0.42 seconds distinguishes our method. This rapid inverse design tool has immediate applications in exploring extreme metamaterials, enabling quick and efficient generation of metamaterial structures.  This approach also opens up possibilities for sequence interpolation in metamaterials, paving the way for designing materials with tailored mechanical responses. Additionally, the tool facilitates the generation of diverse microstructures, enhancing multiscale design considerations.

Our generative tool, with its rapid design capabilities, holds immense value not only in structural engineering but also in other mechanical systems. Reducing the number of optimization cycles decreases the time and resource consumption, enhances the research efficiency, and renders metamaterial development more cost-effective. Its potential impact extends to the biomedical domain, where it can significantly shorten the customization cycle for bone implants, benefiting patients. In conclusion, our proposed fast inverse design method, driven by an advanced deep generative AI algorithm, significantly advances the field of mechanical metamaterials and has the potential to inspire and drive future research endeavors in a wide range of fields, including materials science and engineering.

\section*{Methods}

\subsection*{Data generation}
\paragraph{Data format}
Density-based microstructures are obtained by the inverse homogenization method in a cube domain. The cube is discretized into $n$ voxel elements, with each element assigned a density value of $0$ (void) or $1$ (solid). 
Various combinations of $0$ and $1$ give rise to different microstructures, resulting in $2^n$ combinations. 
The number of microstructure variations expands exponentially with the resolution parameter $n$.  Consequently, voxel-based representations exhibit significant representational capabilities. However, it is important to note that not all combinations yield satisfactory microstructures, and the large amount of data may impose a considerable burden on subsequent training and machine learning algorithms. 
When constructing the dataset, we consider the cube symmetry employing $\frac{1}{8}$ of unit cell lattice to simplify the representation.
Subsequently, symmetric operations are applied to the $\frac{1}{8}$ unit cell lattice. This process ultimately leads to a cubic symmetric microstructure to facilitate calculation. 
To create a sufficiently large design space for a metamaterial database based on voxels, we establish microstructures at a resolution of $128^3$ and subsequently store $\frac{1}{8}$-th of it, equivalent to $64^3$ microstructures.

\paragraph{Data statistics}
The LIVE3D framework~\cite{zhang2023optimized} is used to generate the initial dataset. The objective functions are the bulk modulus, shear modulus, and Poisson's ratio with volume constraints.  
The total optimization results consist of 5,000 structures each for bulk modulus and shear modulus, while 4,396 structures are obtained targeting Poisson's ratio. 
The volume fractions considered in the optimization range from 0.05 to 1.
After iterative augmentation via active learning, our final dataset comprises 144,054 microstructures, as illustrated in Figure~\ref{fig:dataset}. 
However, the distribution of these 144,054 data points in the C11-C12-C44-vol space is uneven. To eliminate bias and improve training efficiency, we remove some data from overly dense regions, resulting in a training set containing 21,212 data points.

\subsection*{Framework of the diffusion model}
The backbone denoising network is a U-Net based on the standard 3D convolutional neural network. Following LAS-Diffusion~\cite{zheng2023locally}, U-Net is composed of 5 levels: $64^3, 32^3, 16^3, 8^3$, and $4^3$, with feature dimensions gradually increasing to 32, 64, 128, 256, and 256, respectively. Each level is constructed with a ResNet block, which includes two convolution layers with a kernel size of 3. In the U-Net bottleneck, two ResNet blocks are introduced. To map the voxel features at the finest level to a surface occupancy value, a convolution layer is added at the conclusion of the network.

\subsection*{Universal guidance for interpolation}
Specifically, universal guidance suggests that the classifier in classifier guidance can be replaced with any objective function, and its gradient can be used to guide the generation. 
The objective function we use for Figure~\ref{fig:interpolation}c is:
$$E(\hat{x}_{0,i}) = -3 f(p(\hat{x}_{0,i}))+  E_{boundary}(p(\hat{x}_{0,i}), x_{0,i-1}) + \alpha  E_{boundary}(p(\hat{x}_{0,i}), x_{0,start}) +(1-\alpha) E_{boundary}(p(\hat{x}_{0,i}), x_{0,end}). $$
Here, $\hat{x}_{0,i}$ is the predicted clear microstructure in the reverse denoising process for the ith microstructure in the interpolation sequence. $p(x) = \frac{tanh(64x+1)}{2}$ is a projection function to force $x$ to a 0-1 tensor. $f$ is a regressor that predicts the ratio of the bulk modulus of the microstructure to its upper limit at the volume fraction. 
The regression-derived $R^2$ is 0.99. 
$E_{boundary}(x,y) = ||x[0,:,:]-y[0,:,:]||_2$, which measures the boundary difference between two microstructures $x$ and $y$. 
The last three terms of the objective function correspond to the boundary differences between the currently predicted microstructure in the denoising process and the previous microstructure in the interpolation sequence, the initial microstructure, and the final microstructure. $\alpha = \frac{i}{N}$, where $N$ is the length of the interpolation sequence.

\subsection*{Multiscale design}
Our multiscale optimization method is an almost trivial extension of the 2D case~\cite{wang2022mechanical}. 
A key difference is that they use microstructures from the dataset to fill the design space, whereas we use the generated microstructures. 
Therefore, similar to their use of property space constraints to restrict the design variables within the dataset range, we also need constraints to control the design variables within the generable range. We consider the generable range to be the coverage of the final version of the dataset. 
Specifically, for every element $e$ in the design space, there is a constraint: $\varphi(\mathbf{C}^e)\geq 0$ to constrain the elasticity tensor $\mathbf{C}$ within the coverage of the final version of the dataset. 
This results in $40\times40\times40$ constraints. As suggested by Wang et al.~\cite{wang2022mechanical}, a single constraint is used to approximate these constraints. However, since our problem scale is much larger than that of their 2D problem, this approximation becomes less accurate, and the $40\times40\times40$ constraints are not always fully satisfied.
Therefore, at the end of each optimization step, we project the elements that fall outside the dataset coverage to their best approximations within the dataset coverage. This ensures that all the elements can be accurately generated.


\section*{Data availability}
The training data, which include voxel-based microstructures and their effective homogenized properties, as well as the generated data optimized to approach the extreme and the interpolated microstructure sequences generated in this study, have been deposited in \href{https://drive.google.com/drive/folders/1fNj_v-8YjtYCPoyXn6qZ-HzG0LqAJeV9}{Google Drive}. 

\section*{Code availability}
The code used to train the generative modeling framework and obtain the inverse design of voxel-based microstructures has been uploaded to  \href{https://github.com/yyy1336/microstructure_generation_3d}{GitHub}.

\bibliography{sample}

\section*{Additional information}
\textbf{Competing interests:} The authors declare that they have no competing interests.

\section*{List of Supplementary Materials}
\begin{itemize}
    \item Supplementary detailed description of dataset generation
    \item Supplementary description of training details
    \item Supplementary description of diffusion model performance
    \item Supplementary description of gradient-based optimization to approach extreme
    \item Supplementary description of the regressor in microstructure interpolation
    \item Captions for Movie 1 to 5
\end{itemize}








\end{document}